**Light Gradient Boosting Machine as a Regression Method for Quantitative Structure-Activity Relationships**

Robert P. Sheridan [1], Andy Liaw, [2] Matthew Tudor [3]

1. Computational and Structural Chemistry, Merck & Co. Inc., Kenilworth, NJ, U.S.A. 07033
2. Biometrics Research Department, Merck & Co. Inc., Rahway, NJ, U.S.A. 07065
3. Computational and Structural Chemistry, Merck & Co. Inc., West Point, PA, U.S.A 19486

sheridan@merck.com

ORCHID

| | |
|---|---|
| Sheridan | 0000-0002-6549-1635 |
| Liaw | 0000-0002-2233-7427 |
| Tudor | 0000-0001-8766-195X |

ABSTRACT

In the pharmaceutical industry, where it is common to generate many QSAR models with large numbers of molecules and descriptors, the best QSAR methods are those that can generate the most accurate predictions but that are also insensitive to hyperparameters and are computationally efficient. Here we compare Light Gradient Boosting Machine (LightGBM) to random forest, single-task deep neural nets, and Extreme Gradient Boosting (XGBoost) on 30 in-house data sets. While any boosting algorithm has many adjustable hyperparameters, we can define a set of standard hyperparameters at which LightGBM makes predictions about as accurate as single-task deep neural nets, but is a factor of 1000-fold faster than random forest and ~4-fold faster than XGBoost in terms of total computational time for the largest models. Another very useful feature of LightGBM is that it includes a native method for estimating prediction intervals.

## INTRODUCTION

Quantitative Structure-Activity Relationships (QSAR) models are very useful in the pharmaceutical industry for predicting on-target and off-target activities. While higher prediction accuracy is desirable, computational efficiency is also important. In an industrial environment there may be a dozens of models trained on a very large number (>100,000) of compounds and a large number (several thousand) of substructure descriptors. These models may need to be updated frequently (say, every few weeks). QSAR methods that are particularly compute-intensive or require the adjustment of many sensitive hyperparameters to achieve good prediction for an individual QSAR data set are therefore less desirable. Other useful characteristics of a good method include the ability to make predictions rapidly, produce interpretable models, and indicate how reliable each prediction might be.

Random forest (RF) (Breiman, 2001; Svetnick et al., 2003) was attractive as a QSAR method for many years because it is easily parallelizable and has few adjustable hyperparameters, as well as being relatively accurate in prediction. The most notable recent trend in the QSAR literature is the use of deep neural nets (DNN) (Gawehn et al., 2016), starting with our original work stemming from our Kaggle competition (Ma et al, 2015). On the plus side, DNNs tend to make very accurate predictions. On the other hand, DNNs tend to be computationally expensive and their models are not easily interpretable. Gradient boosting is another interesting QSAR method that has undergone rapid development. We investigated Extreme Gradient Boosting (XGBoost) by Chen and Guestrin (2016) on QSAR problems (Sheridan et al., 2016). Later we examined BART (Feng et al., 2019). While boosting methods have a very large number of adjustable hyperparameters, we can show that, in certain ranges, predictivity is at a near maximum and one can pick a standard set of hyperparameters for XGBoost with which most QSAR problems can achieve a level of prediction about as good as DNN, with much less computational expense.

Recently a new method of gradient boosting called Light Gradient Boosting (LightGBM) has appeared (Ke et al., 2017). LightGBM differs from XGBoost in a few aspects. We will only discuss the ones that are relevant in the context of QSAR. In XGBoost, trees are grown to a pre-specified depth; i.e., it will not split nodes to the k+1-st level until it had performed all possible splits at the k-th level. LightGBM, on the other hand, will split the node that maximizes the drop in the loss function (thus it can grow "lop-sided" trees). This is the "best-first" strategy of regression tree induction described in Friedman (2001). (This feature has also recently been added to the XGBoost software as an option.) In addition, LightGBM introduced a feature, "exclusive feature bundling", which collapse sparse descriptors (those with very few non-zero elements, and the non-zero elements do not occur in the same molecules) into one feature. This not only increase the computational efficiency, but can increase the information content of descriptors seen by the algorithm. There are other features in LightGBM that increase the computational efficiency in terms of time and memory usage.

One feature of LightGBM that was not in the original formulation of XGBoost is a method for assigning uncertainty of predictions. Uncertainty of predictions can be estimated via prediction intervals; i.e., the interval (L, U) should have 95% chance of containing the measured activity. The wider the interval (U – L), the higher the uncertainty in the prediction. LightGBM can be used to estimate these intervals by using a quantile loss function.

Several groups have compared boosting methods on a number of machine learning applications. The claim for general machine learning problems is that LightGBM is much faster than XGBoost and takes less memory (Omar, 2017; Anghel et al. 2019; Du et al., 2019). A recent paper by Zhang et al. (2019) applies LightGBM to classification problems, specifically to toxicity and compares its performance to RF, DNN, and XGBoost in random cross-validation. This paper compares LightGBM against RF, DNN, and XGBoost as a regression method for prospective prediction on a wider variety of QSAR problems. We can show that a subset of hyperparameters can be found at which LightGBM is faster than XGBoost and achieves prediction accuracies equivalent to single-task DNN. We also examine the prediction intervals from LightGBM in comparison to RF and BART.

## METHODS

### Data sets

Table 1 shows the in-house data sets used in this study which are the same as in Ma et al. (2015) and Sheridan et al. (2016). These data sets represent a pharmaceutical research relevant mixture of on-target and off-target activities, easy and hard to predict activities, and large and small data sets. Descriptors for the data sets (in disguised form) are available in Supporting Information of Sheridan et al. (2020). As before, we use in-house data sets because:
1. We wanted data sets which are realistically large, and whose compound activity measurements have a realistic amount of experimental uncertainty and include a non-negligible amount of qualified data.
2. Time-split validation (see below), which we consider more realistic than any random cross-validation, requires dates of testing, and these are very hard to find in public domain data sets.

Table 1. Data sets for prospective prediction.

| Data set | Type | Description | Number of Molecules Training+test | Number of unique AP,DP descriptors | Mean ± stdev activity |
|---|---|---|---|---|---|
| 2C8 | ADME | CYP P450 2C8 inhibition -log(IC50) M | 29958 | 8217 | 4.88±0.66 |
| 2C9BIG | ADME | CYP P450 2C9 inhibition -log(IC50) M | 189670 | 11730 | 4.77±0.59 |
| 2D6 | ADME | CYP P450 2D6 inhibition -log(IC50) M | 50000 | 9729 | 4.50±0.46 |
| 3A4* | ADME | CYP P450 3A4 inhibition | 50000 | 9491 | 4.69±0.65 |
| | | -log(IC50) M | | | |
| A-II | Target | Binding to Angiotensin-II receptor -log(IC50) M | 2763 | 5242 | 8.70±2.72 |
| BACE | Target | Inhibition of beta-secretase -log(IC50) M | 17469 | 6200 | 6.07±1.40 |
| CAV | ADME | Inhibition of Cav1.2 ion channel | 50000 | 8959 | 4.93±0.45 |
| CB1* | Target | Binding to cannabinoid receptor 1 -log(IC50) M | 11640 | 5877 | 7.13±1.21 |
| CLINT | ADME | Clearance by human microsome log(clearance) μL/min/mg | 23292 | 6782 | 1.93±0.58 |
| DPP4* | Target | Inhibition of dipeptidyl peptidase 4 -log(IC50) M | 8327 | 5203 | 6.28±1.23 |
| ERK2 | Target | Inhibition of ERK2 kinase -log(IC50) M | 12843 | 6596 | 4.38±0.68 |
| FACTORXIA | Target | Inhibition of factor Xla -log(IC50) M | 9536 | 6136 | 5.49±0.97 |
| FASSIF | ADME | Solubility in simulated gut conditions log(solubility) mol/l | 89531 | 9541 | -4.04±0.39 |
| HERG | ADME | Inhibition of hERG channel -log(IC50) M | 50000 | 9388 | 5.21±0.78 |
| HERGBIG | ADME | Inhibition of hERG ion channel -log(IC50) M | 318795 | 12508 | 5.07±0.75 |
| HIVINT* | Target | Inhibition of HIV integrase in a cell based assay -log(IC50) M | 2421 | 4306 | 6.32±0.56 |
| HIVPROT* | Target | Inhibition of HIV protease -log(IC50) M | 4311 | 6274 | 7.30±1.71 |
| LOGD* | ADME | logD measured by HPLC method | 50000 | 8921 | 2.70±1.17 |
| METAB* | ADME | percent remaining after 30 min microsomal incubation | 2092 | 4595 | 23.2+/-33.9 |
| NAV | ADME | Inhibition of Nav1.5 ion channel -log(IC50) M | 50000 | 8302 | 4.77±0.40 |
| NK1* | Target | Inhibition of neurokinin1 (substance P) receptor binding -log(IC50) M | 13482 | 5803 | 8.28±1.21 |
| OX1* | Target | Inhibition of orexin 1 receptor -log(Ki) M | 7135 | 4730 | 6.16±1.22 |
| OX2* | Target | Inhibition of orexin 2 receptor -log(Ki) M | 14875 | 5790 | 7.25±1.46 |
| PAPP | ADME | Apparent passive permeability in PK1 cells log(permeability) cm/sec | 30938 | 7713 | 1.35±0.39 |
| PGP* | ADME | Transport by p-glycoprotein log(BA/AB) | 8603 | 5135 | 0.27±0.53 |
| PPB* | ADME | human plasma protein binding log(bound/unbound) | 11622 | 5470 | 1.51±0.89 |
| PXR | ADME | Induction of 3A4 by pregnane X receptor; percentage relative to rifampicin | 50000 | 9282 | 42.5±42.1 |
| RAT_F* | ADME | log(rat bioavailability) at 2mg/kg | 7821 | 5698 | 1.43±0.76 |
| TDI* | ADME | time dependent 3A4 inhibitions log(IC50 without NADPH/ IC50 with NADPH) | 5559 | 5945 | 0.37±0.48 |
| THROMBIN* | Target | human thrombin inhibition -log(IC50) M | 6924 | 5552 | 6.67±2.02 |

* Kaggle data sets

As an example of qualified data, one might know that the measured IC50 > 30μM only because 30μM was the highest concentration titrated, and the assay did not reach an inflection point up to that dose. For the purposes of model-building those activities are treated as fixed numbers, because most off-the-shelf QSAR methods handle only fixed numbers. For example, IC50 > 30μM is set to IC50=30 X $10^{-6}$ M or –log(IC50)=4.5. Our experience is that it is best to keep such qualified data in QSAR training sets; otherwise less active compounds are often predicted to be more active than they really are.

In order to compare the predictive ability of QSAR methods, each of these data sets was split into two non-overlapping subsets: a training set and a test set. Our training and test sets are generated by "time-split". For each data set, the first 75% of the molecules assayed form the training set, while the remaining 25% of the compounds assayed later form the test set. We have found that, for regressions, $R^2$ from time-split validation better estimates the $R^2$ for true prospective prediction than $R^2$ from any "split at random" scheme (Sheridan, 2013). Since training and test sets are not selected from the same pool of compounds, the descriptor distributions in these two subsets are frequently not the same.

## QSAR Descriptors

Each molecule is represented by a list of features, i.e. "descriptors" in QSAR nomenclature. In this paper, we use a set of descriptors that is the union of AP, the original "atom pair" descriptor from Carhart et al. (1985). and DP descriptors ("Donor acceptor Pair"), called "BP" in Kearsley et al. (1996) Both descriptors are of the form:

Atom type *i* – (distance in bonds) – Atom type *j*

For AP, atom type includes the element, number of nonhydrogen neighbors, and number of pi electrons; it is very specific. For DP, atom type is one of seven (cation, anion, neutral donor, neutral acceptor, polar, hydrophobe, and other); it contains a more generic description of chemistry.

## Random Forest

RF is an ensemble recursive partitioning method where each recursive partitioning "tree" is generated from a bootstrapped sample of compounds, and a random subset of descriptors is used at the branching of each node in the tree. While there are a few adjustable hyperparameters (e.g. number of trees, fraction of descriptors used at each branching, and size of nodes below which no further splitting should be done), the quality of predictions is generally insensitive to changes in these hyperparameters.

The version of RF we are using is a modification of the original FORTRAN code from Breiman (2001), which is built for regressions. It has been parallelized to run one tree per processor on a cluster. Such parallelization is necessary to run some of our larger data sets in a reasonable time. For all RF models, we generate 100 trees with m/3 descriptors used at each branch-point, where m is the number of unique descriptors in the training set. The tree nodes with 5 or fewer molecules are not split further. We apply these hyperparameters to every data set.

### Deep Neural Nets

Our Python-based DNN code for fully-connected neural nets is the one obtained through the Kaggle competition from George Dahl (Dahl et al., 2014), then at the U. of Toronto, and modified by us, and deposited in GitHub (https://github.com/Merck/DeepNeuralNet-QSAR). The DNN results we present are for single-task regression models using the "standard parameters" in Ma et al. (2015), which are applied to all data sets. For timing purposes, we also have implemented a simplified ("quick") version of the DNN, which achieves almost identical prediction accuracy to the standard parameters, but uses a smaller neural net. Those parameters are in Sheridan et al. (2016).

### XGBoost

The implementation of XGBoost is the C++ version runnable on Linux. https://picnet.com.au/blog/xgboost-windows-x64-binaries-for-download/ . There are several dozen adjustable hyperparameters of which four we consider "standard" for QSAR problems. These are given in Sheridan et al. (2016). We were able to show that these standard parameters, which are used for all datasets, lead to predictions as good as those where the parameters were optimized for each dataset separately.

### Light Gradient Boosting Machine

We are using the Python version downloadable from https://github.com/microsoft/LightGBM. The version used for the current study is 2.2.2.

### Metrics

In this paper, the metric used to evaluate prediction performance is $R^2$, the squared Pearson correlation coefficient between predicted and observed activities in the test set. $R^2$ is an attractive measurement for model comparison across many data sets because it is unitless and ranges from 0 to 1 for all data sets. The relative predictivity of the three methods we examine does not change if we use alternative metrics such as $Q^2$ or RMSE.

Workflow for LightGBM hyperparameter optimization

One goal is to identify a set of hyperparameters that would be useful for most QSAR data sets, as was done for other methods. We considered 5 hyperparameters for optimization, the rest were the default.

*nrounds*, the total number of trees in the model

*learnrate*, the weight on each tree

*nleaves*, the maximum number of leaves per tree. An alternative to controlling the complexity of an individual tree is *maxdepth*, which is the maximum depth of a tree. The maximum possible number of leaves per tree is $2^{maxdepth}$.

*bagfrac*, the fraction of compounds randomly selected to create each tree

*featfrac*, the fraction of descriptors randomly selected to create each tree

We follow a similar workflow for hyperparameter optimization as we did with XGBoost (Sheridan et al., 2016). Whereas for XGBoost we optimized three hyperparameters in a full grid search, for five hyperparameters a step-wise grid search was more practical. For each dataset the hyperparameters were optimized in the following way.

1. An "original" set of hyperparameter was *nrounds*=1500, *learnrate*=0.01, *nleaves*=32, *bagfrac*=0.7, *featfrac*=0.7.
2. A grid search was done on *nrounds*=(1500,700,350,100) and *learnrate*=(0.01,0.02,0.05,0.1). Other hyperparameters as in the original.
3. Given the optimum combination of *nrounds* and *learnrate*, a grid search was done on *nleaves*=(16,32,64,128,256).
4. Given the optimum combination of *nrounds*, *learnrate*, and *nleaves*, a grid search was done for *featfrac*=(0.25,0.50,0.7,1.0) and *bagfrac*=(0.25,0.50,0.7,1.0)

Searches were done under two different circumstances.

1. TESTOPT: Find the optimum combination of hyperparameters that gives the highest average $R^2$ for the test sets. It should be noted that TESTOPT does not reflect a realistic situation, because we would not know the activity values of the test set in advance. However, this gives us an upper limit for the $R^2$ on the test set we can expect by optimizing these hyperparameters, and it is interesting to know what hyperparameter values we should use *if* we had prior knowledge. TESTOPT finds a different set of optimum hyperparameters for each data set.

2. TRAINOPT: Finding the optimal values by cross-validation of each training set. That is, split the training set in half randomly, make a model from the first half using the hyperparameters, and then predict the remaining half. The set of hyperparameters that gives the highest $R^2$ for the prediction of the second half of the training set is used to generate a model using the entire training set. This model is used to predict the test set. This is more realistic situation because we are optimizing only on the training set. TRAINOPT finds a different set of hyperparameters for each data set.

3. STANDARD. The goal is to find a common value of *nrounds*, *learnrate*, *nleaves*, *bagfrac*, and *featfrac* to be used for all data sets. The most straightforward way of generating such a standard set is to find the mean optimum values of the five hyperparameters in TESTOPT, TRAINOPT, or both.

Assessing Uncertainty of Predictions

A way to assess uncertainty of prediction is with prediction intervals: A 95% prediction interval (L,U) for a molecule indicates that with 95% probability the prediction on the model derived from the actual data should be larger than L and smaller than U. One typical method of generating these intervals is via quantile regression. For RF, the algorithm can be exploited to generate such intervals with some manipulation of output (Meinshausen, 2006). When predicting the activity of a molecule, for each tree in the forest, the molecule will end up in a terminal node. The molecules in the training set that also landed in that terminal nodes are considered neighbors of the molecules being predicted. Aggregating these neighbors across all trees in the forest, we can form prediction intervals by computing the weighted $2.5^{th}$ and $97.5^{th}$ weighted percentiles of the activities of these neighbors, the weights being the frequency that a molecule appears as a neighbor over all trees. BART, being a Bayesian method, naturally outputs a distribution for each point being predicted, and the quantiles of the distribution (e.g., $2.5^{th}$ and $97.5^{th}$ percentiles) serve as the prediction interval. LightGBM provides the option of the quantile loss function that can be used to predict the given quantile. To generate prediction intervals, one would build two LightGBM models, one for the lower limit (e.g., the $2.5^{th}$ percentile) and another for the upper limits (e.g., $97.5^{th}$ percentile).

Timing

The three methods we are comparing RF, DNN, and XGBoost/LightGBM work on different machine architectures and/or modes in our environment:

1. RF runs as 100 jobs (one for each tree) running in parallel on a cluster. The cluster nodes are HP ProLiant BL460c Gen8 server blades, each equipped with two 8 core Intel(R) Xeon(R) CPU E5-2670 0 @ 2.60GHz processors and 256GB Random Access Memory (RAM). The total time is the time for a single job times 100.

2. XGBoost and LightGBM run on a single node of the above cluster.

3. DNN runs on a single NVIDIA Tesla C2070 GPU . The GPU runs almost exactly 10-fold faster than the cluster nodes, so the total time for a DNN on a cluster node would be 10 times the time for running on the GPU.

Model size

Another interesting aspect of a QSAR method is the size of the model file, in that the speed of prediction is sometimes limited in practice by the time taken to read the model into memory or by copying the model from an archive to the computer doing the predicting. Here we note the size of the (binary) files comprising the model. In the case of RF, we multiply the size of a single tree by 100. In practice the size of a model file can vary depending on the particular implementation of the QSAR method, including compression of the file, but looking at the size will give us a rough idea of the relative complexity of models from the respective methods.

## RESULTS

### Optimizations and standard hyperparameters for LightGBM

The optimum set of hyperparameters for individual data sets and the $R^2$ for each type of optimization are in Supporting Information. As with XGBoost, we find that the prediction accuracy is not particularly sensitive to the hyperparameters we examined. Mean optimum hyperparameter values are shown in Table 2. While the optimum values of hyperparameters do not correlate well between TESTOPT and TRAINOPT for individual data sets (not shown), the mean optimum hyperparameter values for TESTOPT and TRAINOPT are not far apart relative to the overall range of each hyperparameter. We averaged over both TESTOPT and TRAINOPT for all datasets to obtain the standard hyperparameters. It is interesting that the STANDARD value of *bagfrac* is close to the fraction of compounds that would appear through bagging in random forest (0.66), and the STANDARD value of *featfrac* is close to the descriptors/3 (0.33) rule for regressions in random forest. Predictions using these values are called LightGBM_STANDARD. An independent study from our laboratory (DiFranzo et al., 2020) used the same datasets, but built models using only compounds similar to those in the test set. The hyperparameters for LightGBM, generated by a more automated procedure, are also listed in Table 2. They are not dissimilar.

Table 2. Mean optimal values for five hyperparameters averaged over 30 datasets

| Hyperparameter | TESTOPT | TRAINOPT | STANDARD | DiFranzo et al. |
|---|---|---|---|---|
| *nrounds* | 1089 ± 550 | 1306±342 | 1200 | 1000 |
| *learnrate* | 0.027 ± 0.019 | 0.029±0.02 | 0.028 | 0.05 |
| *nleaves* | 84 ± 77 | 58±51 | 72 | ≤128 (*maxdepth*=7) |
| *bagfrac* | 0.68 ± 0.24 | 0.73±0.26 | 0.71 | 0.5 |
| *featfrac* | 0.45 ± 0.25 | 0.35±0.19 | 0.40 | 0.2 |

In TRAINOPT there is a relationship between the optimum *learnrate*, *nleaves* and *bagfrac* vs. Ntraining, the number of molecules in the training set: smaller data sets tend to prefer smaller values of these hyperparameters. Therefore, it might be possible to guess a good value for these hyperparameters for individual data sets based only on Ntraining. However, we will show below that the STANDARD hyperparameters already give almost as good predictions as the TRAINOPT grid-search, so this type of refinement is not likely to be helpful overall. We made similar observations with XGBoost.

### Accuracy of prediction for the QSAR methods

Figure 1 (Top) shows the $R^2$ for prediction of the test set for DNN_STANDARD, XGBOOST_STANDARD, LightGBM_TRAINOPT and LightGBM_STANDARD vs. the $R^2$ for RF, which we are taking as the baseline method. Figure 1 (Bottom) shows the $R^2$ for prediction minus the $R^2$ for RF vs. the $R^2$ for RF, which better shows small differences between methods. Generally speaking, LightGBM_TRAINOPT might be more predictive than LightGBM_STANDARD, indicating there

might theoretically be a reason to optimize the hyperparameters for individual datasets, but this difference is very small; both are generally at least as good as DNN in predictivity. Thus, in our opinion, the much greater time to optimize QSAR datasets would not be justified, and STANDARD hyperparameters could be used to good effect. The overall predictivity of LightGBM_STANDARD hyperparameters is very close to that of the original hyperparameters, supporting the idea that a range of hyperparameter values is acceptable.

The visual impression in Figure 1 is consistent with the mean $R^2$ over the 30 data sets as shown in Table 3. Although representing an unrealistic scenario, one would expect LightGBM_TESTOPT set results to get slightly higher predictivity than LightGBM_TRAINOPT, and it does. These are only averages; any of the four methods may do best on a particular data set.

Table 3. Mean $R^2$ for 30 datasets for methods using the AP,DP descriptor:

| Method | Mean $R^2$ |
|---|---|
| RF | 0.39 |
| DNN_STANDARD | 0.43 |
| XGBoost_STANDARD | 0.43 |
| LightGBM_TRAINOPT | 0.45 |
| LightGBM_STANDARD | 0.44 |
| LightGBM_TESTOPT | 0.46 |

Timing

Total compute time is shown in Figure 2 as a function of Ntraining. The log-log plot is the one where all methods show a maximally linear correlations and the range of timings can be appreciated. The total compute effort for DNN, XGBoost, and LightGBM using standard hyperparameters are roughly linear with Ntraining. As expected, the DNN using fewer neurons ("quick") requires less computation than the standard DNN. The total compute effort for RF rises roughly as the square of Ntraining. Clearly, boosting methods are much faster than RF and DNN in total compute effort, at least for the larger data sets, and LightGBM is faster than XGBoost by a factor of almost 4.

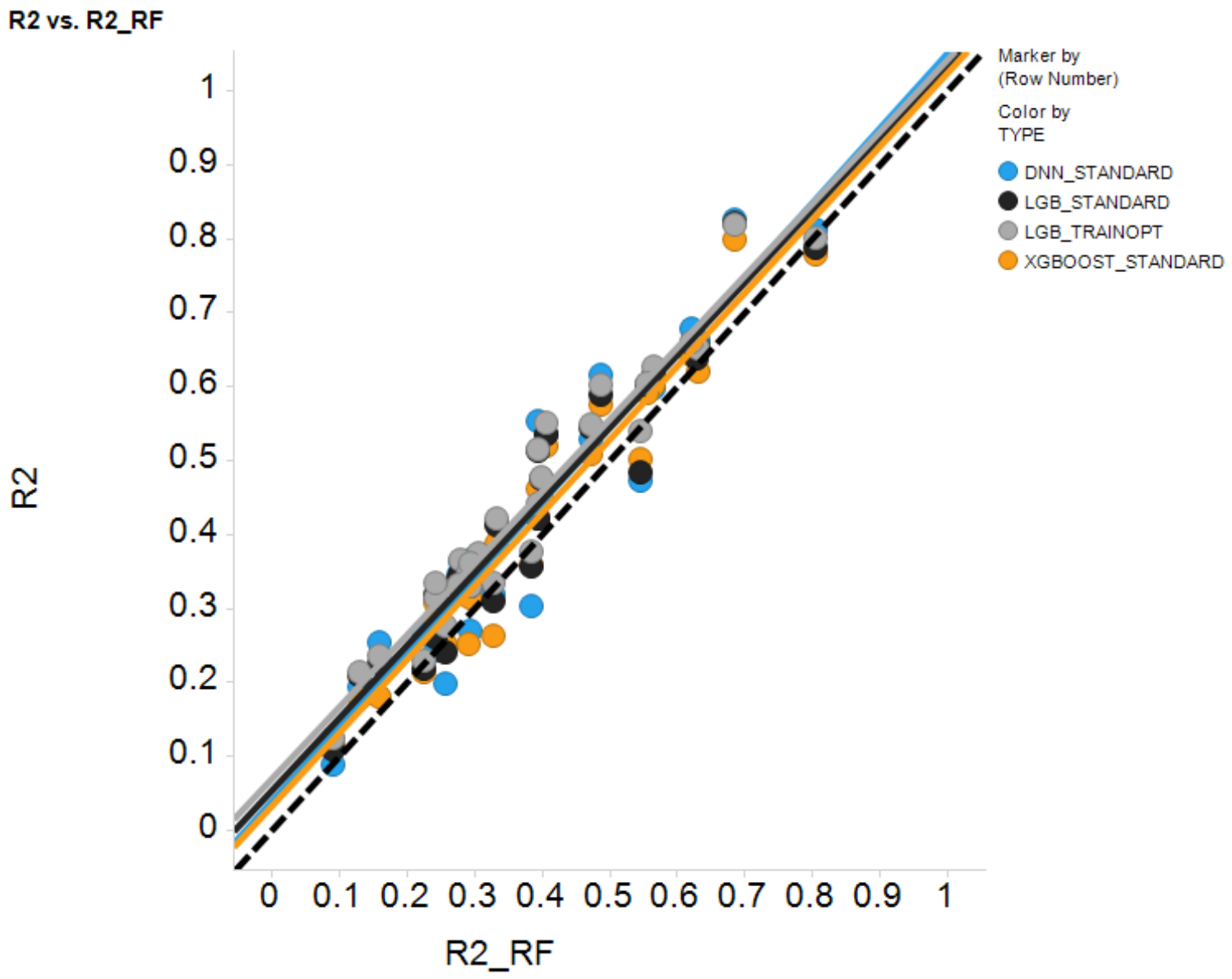

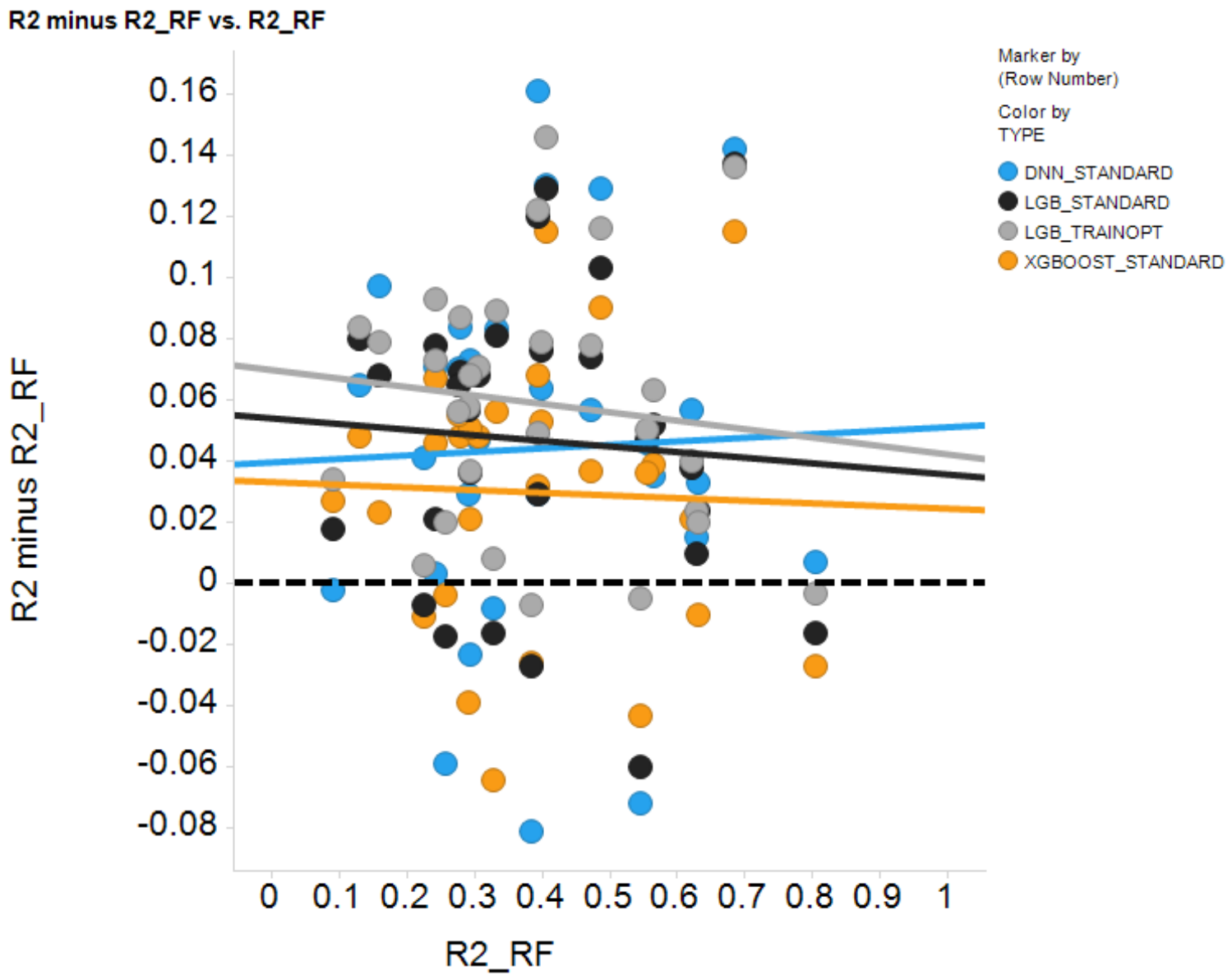

Figure 1. Prediction accuracy on the test set for DNN, XGBoost, and LightGBM and deep neural nets vs. the prediction accuracy of random forest. Two different types of LightGBM hyperparameters are shown, one with the hyperparameters optimized for individual training sets (grey), and one using a standard set of hyperparameters for all data sets (black). (Top) The absolute $R^2$. (Bottom) The $R^2$ minus the $R^2$ for random forest.

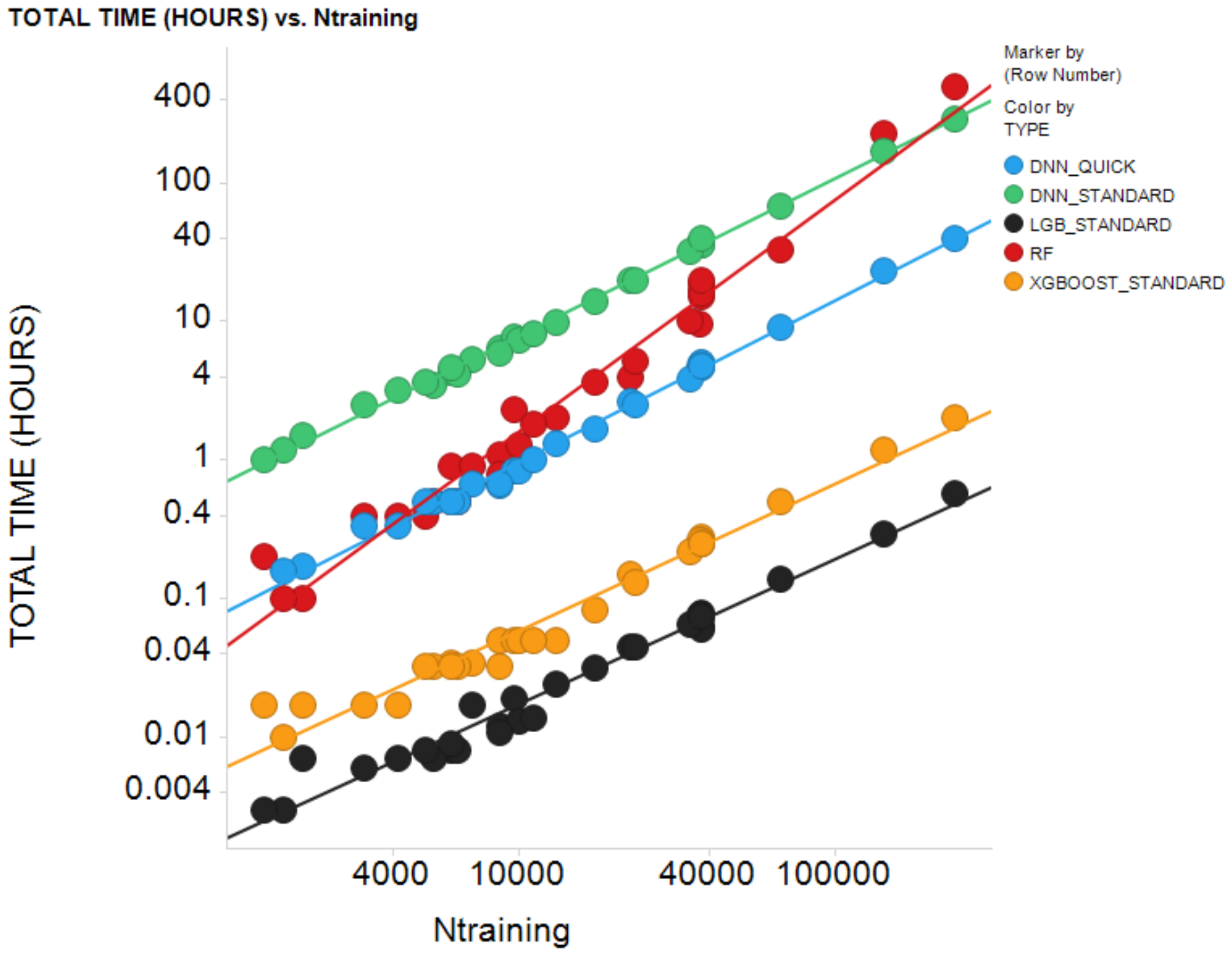

Figure 2. Total computational effort (Top) for random forest, deep neural nets, XGBoost, and LightGBM. The computational effort is expressed in units of hours on a single cluster node.

Model size

Total model file size (in megabytes) is shown in Figure 3 as a function of Ntraining. The log-log plot is the one where all methods show a maximally linear correlations and the range of model file sizes can be appreciated. The size of DNN models is expected to depend on the total number of neurons. The number of neurons of the lowest layer will depend on the number of descriptors, which varies roughly as log(Ntraining), and the number of neurons in the intermediate layers will depend on the number of intermediate layers and number of neurons per layer set by the user. Effectively, the dependence of size is approximately log(Ntraining). We would expect networks with fewer layers and fewer neurons per layer (the "quick" DNN) to produce smaller models than the original standard DNN, and they do. In contrast, the number of nodes in an unpruned recursive partitioning tree should be linear with Ntraining, and we see this for RF. There is an small dependence of size of XGBoost models roughly with log(Ntraining), which probably reflects the fact that larger data sets have more trees closer to the maximum depth. The model size of LightGBM_STANDARD is constant for all models, at ~7 megabytes, somewhat bigger than for XGBoost. LightGBM models using the original hyperparameters (not shown in the figures) are ~4 megabytes, probably reflecting the smaller value of *nleaves* (32 vs. 72). Both flavors of boosting generate models that are tiny compared to those from RF and DNN.

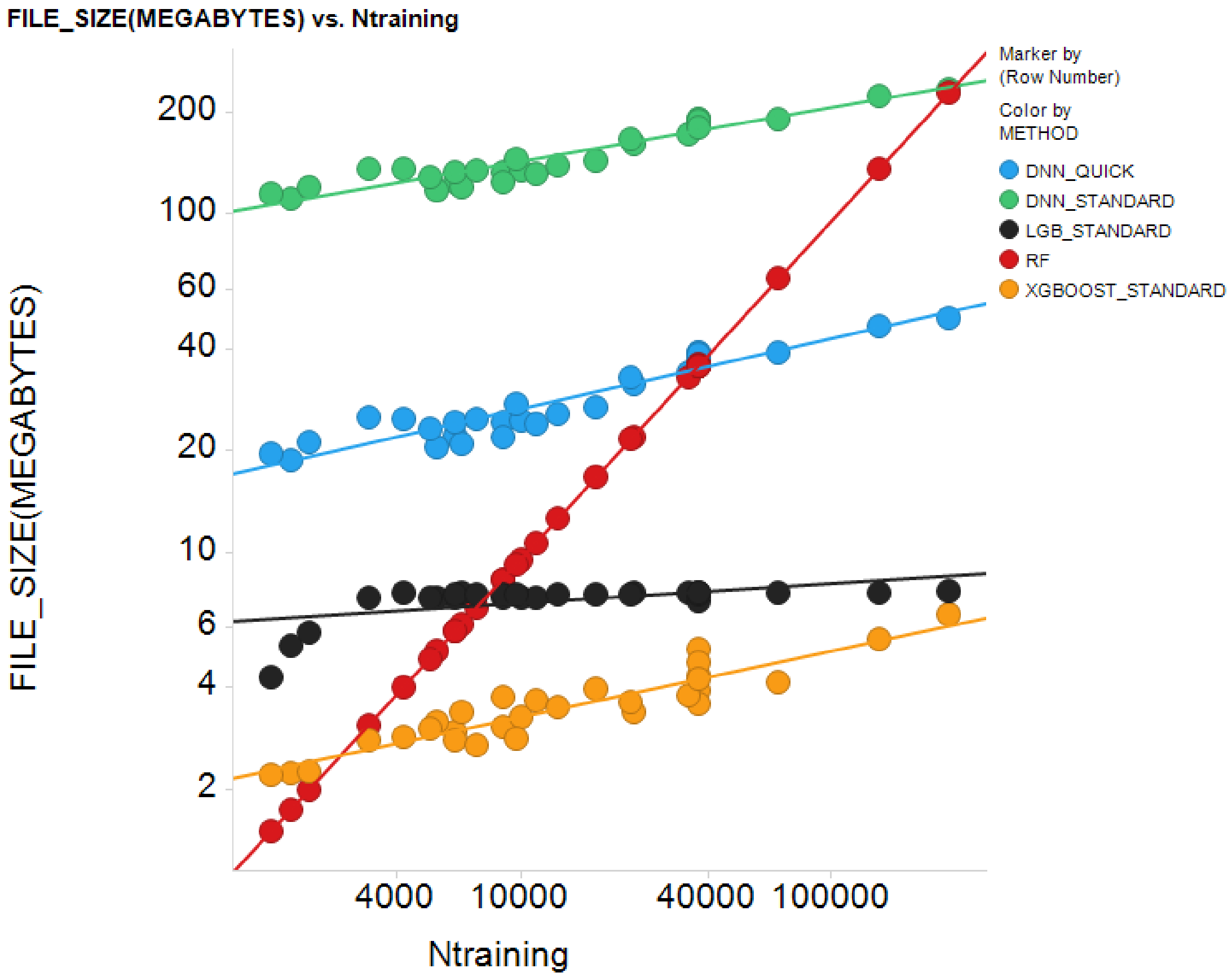

Figure 3. Total model file size for the QSAR methods.

Uncertainty of Prediction

For assessing prediction uncertainty with prediction intervals, we evaluate three methods: BART, quantile regression forest (QRF), and quantile regression using LightGBM (QLGB), by comparing the coverage on the test set data. Coverage is the fraction of the data point for which the observed activity falls within the prediction intervals. Since we generated 95% prediction intervals, we expect to see approximately 95% of the data to fall within ("covered by") the intervals. Figure 4A shows the comparison of the coverage of the three methods on the 30 datasets. Quantile regression using LightGBM gives good coverage for most of the datasets. It does better than quantile regression forests but not as well as BART. Figure 4B shows the median widths of the prediction intervals, normalized to the minimum median width across methods within each data set. For methods with comparable coverages, one naturally prefers methods with smaller prediction intervals (indicating less uncertainty). Even though lightGBM tend to have shorter interval widths compared to BART and RF, one should take into account the coverages are not necessarily comparable.

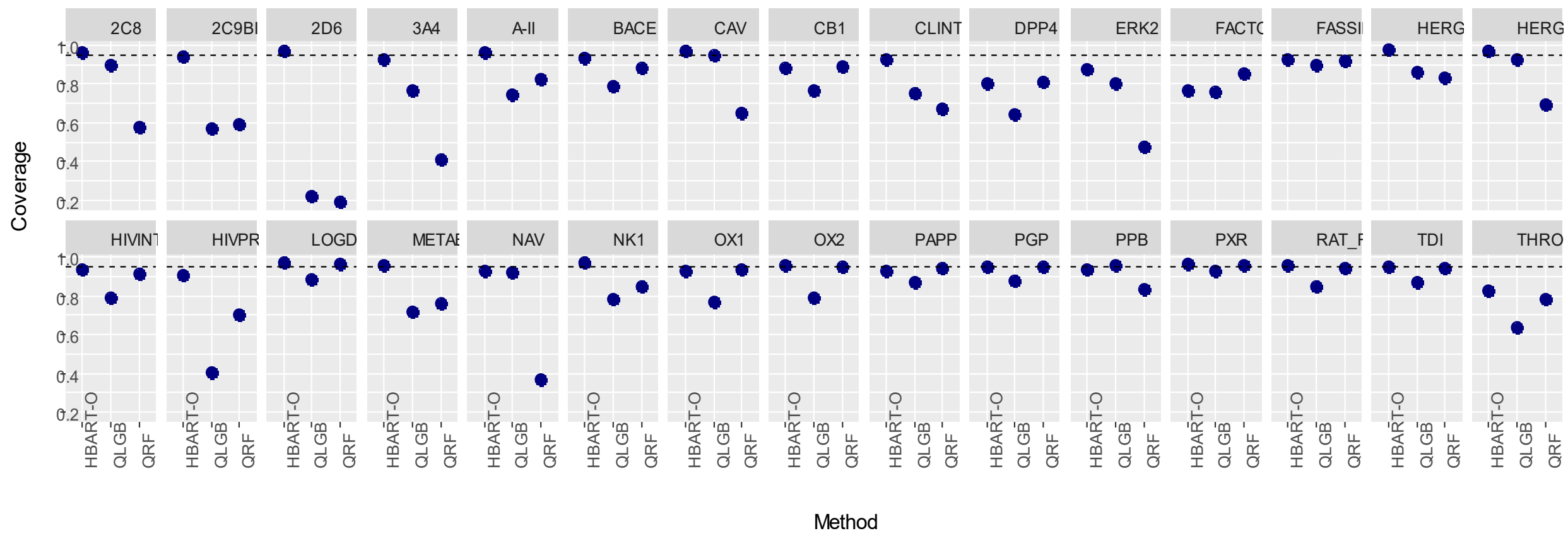

B

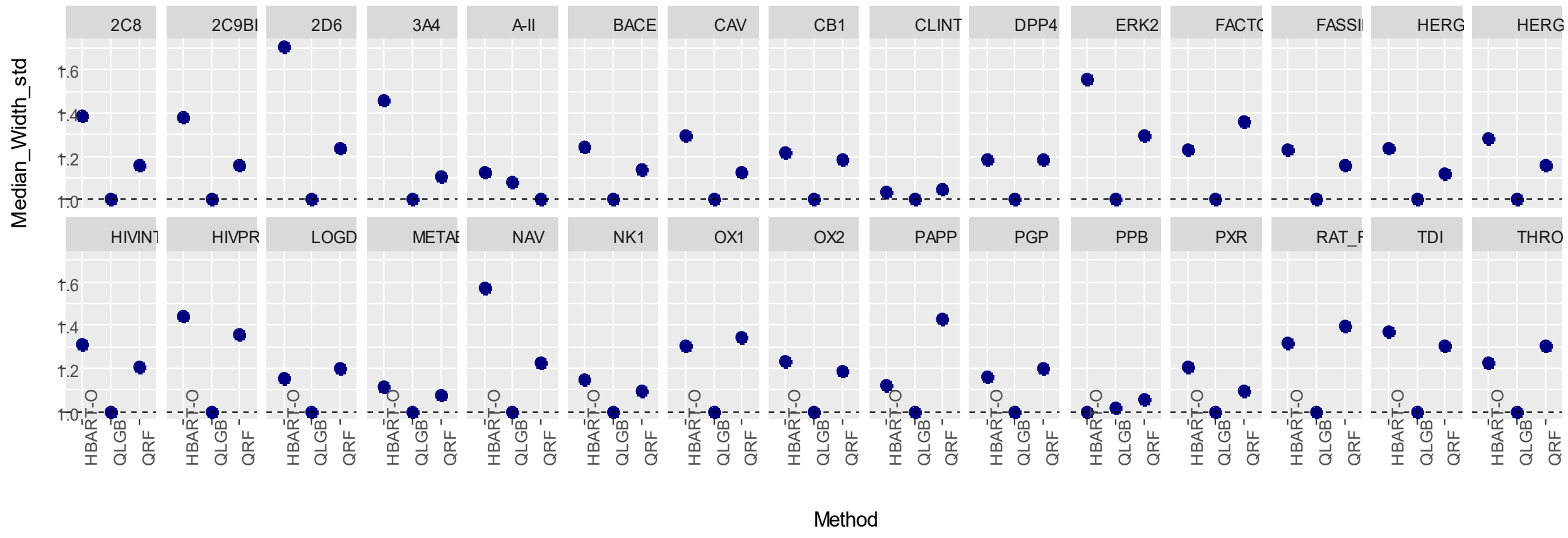

Figure 4. Coverage (A) and median width (B) of the error estimation of BART, Quantile Random Forest, and LightGBM.

## DISCUSSION

Most of the current attention in QSAR is on various deep neural net architectures, and these seem to have an edge over more traditional methods in terms of accuracy in prediction. This is especially true of neural net methods that do not use explicit input descriptors, but use "convolution" on individual atomic or bond properties to effectively generate their own problem-specific descriptor types on-the-fly (Feinberg, et al. 2020; Chuang et al., 2020; Walters and Barzilay, 2021). However, under many circumstances computational efficiency is at least as important as accuracy. Boosting appears to be a very effective and efficient class of machine-learning methods. Previously we showed that XGBoost achieves predictions for QSAR datasets as good as at that from single-task DNN and it does it for orders of magnitude less total compute time and produces much smaller models. Here we demonstrated that LightGBM, tested on diverse QSAR problems, produces slightly better predictions than XGBoost, and takes even less compute time. This is consistent with the observations of Zhang et al. (2019) on toxicity problems using random-split validation. Having LightGBM means we can potentially handle many more and larger data sets and/or update them more frequently than we have previously, given our current compute environment.

The potential difficulty LightGBM having multiple adjustable hyperparameters turns out, in practice, to not be a real issue for QSAR because we can identify standard values of at least some hyperparameters. As we have previously showed with DNN and XGBoost, standard hyperparameters can be used effectively with a large number of QSAR data sets, so that it is not necessary to optimize the hyperparameters for each individual data set. As well as taking a great deal of time, optimization has the additional drawback that it might not be as effective as hoped in true prospective prediction. It is a tacit assumption in QSAR that the molecules to be predicted (in the test set) are similar enough to the training set that maximizing the cross-validated predictions of the training set (by using different descriptors, tweaking adjustable hyperparameters, etc.) is equivalent to maximizing predictivity on the test set. In practice, the training and test sets may be different enough that this is not true. The fact that the optimum hyperparameters in LightGBM_TRAINOPT and LightGBM_TESTOPT do not correlate, and that similar findings were made for XGBoost (Sheridan et al., 2016) and DNN (Ma et al., 2015), supports this.

Recursive partitioning methods like RF, XGBoost, and LightGBM make predictions based on the average observed activities of molecules at their terminal nodes. This has the effect of compressing the range of predictions relative to the observed activities. For random forest we routinely do "prediction rescaling" (Sheridan, 2014), where the self-fit predicted activities in the training set of a particular model are linearly scaled to match the observed activities, and this scaling is applied to further predictions from that model. This does not affect the $R^2$ of prediction, but does help the numerical match of predicted and observed activities at the highest and lowest ranges of activity. We have found XGBoost and LightGBM also benefit from prediction rescaling.

LightGBM also provides a way to assess uncertainty of predictions via quantile regression, something not available in XGBoost. It does entail having to build a separate model for each end of the interval (i.e., one model for the lower limit and another for the upper limit), but the efficiency of the software is such that building extra models with LightGBM can still take less time than with other methods. While LightGBM uncertainties are perhaps not as good in terms of coverage as those in BART (this paper) or as good as Gaussian Processes (DiFranzo et al., 2020), they are somewhat better than those from Quantile RF (this paper). A detailed comparison of the lightGBM intervals to a wider variety of methods will be published elsewhere.

AKNOWLEGEMENTS
We thank Dai Feng for generating the results for BART and quantile random forests in Figure 4.

CONFLICT OF INTEREST

The authors declare no financial conflict of interest.

SUPPORTING INFORMATION

## 1. Hyperparameters for TESTOPT TRAINOPT of LightGBM.

| Data set | nrounds_TESTOPT | learnrate_TESTOPT | nleaves_TESTOPT | featfrac_TESTOPT | bagfrac_TESTOPT | nroonds_TRAINOPT | learnrate_TRAINOPT | nleaves_TRAINOPT | featfrac_TRAINOPT | bagfrac_TRAINOPT |
|---|---|---|---|---|---|---|---|---|---|---|
| 2C8 | 700 | 0.1 | 256 | 0.7 | 0.25 | 1500 | 0.02 | 64 | 0.7 | 0.25 |
| 2C9BIG | 1500 | 0.05 | 256 | 0.5 | 0.25 | 1500 | 0.1 | 256 | 0.7 | 1 |
| 2D6 | 1500 | 0.02 | 128 | 1 | 0.5 | 1500 | 0.02 | 64 | 1 | 0.5 |
| 3A4 | 1500 | 0.05 | 128 | 0.7 | 0.5 | 1500 | 0.05 | 64 | 0.7 | 0.5 |
| ANRINA | 700 | 0.02 | 32 | 1 | 0.7 | 700 | 0.01 | 16 | 1 | 0.25 |
| BACE | 1500 | 0.01 | 32 | 0.5 | 0.5 | 1500 | 0.02 | 32 | 1 | 0.25 |
| CAV | 1500 | 0.02 | 32 | 0.7 | 0.25 | 1500 | 0.05 | 128 | 1 | 0.25 |
| CB1 | 1500 | 0.01 | 64 | 0.5 | 0.7 | 700 | 0.01 | 16 | 0.25 | 0.25 |
| CLINT | 1500 | 0.02 | 128 | 0.5 | 0.5 | 1500 | 0.02 | 64 | 1 | 0.25 |
| DPP4 | 350 | 0.01 | 256 | 0.25 | 0.5 | 1500 | 0.01 | 16 | 0.25 | 0.25 |
| ERK2 | 700 | 0.02 | 16 | 1 | 0.25 | 1500 | 0.01 | 16 | 0.7 | 0.25 |
| FACTORXIA | 100 | 0.01 | 16 | 1 | 1 | 1500 | 0.01 | 32 | 0.7 | 0.25 |
| FASSIF | 1500 | 0.02 | 128 | 0.5 | 1 | 1500 | 0.05 | 128 | 1 | 0.7 |
| HIV_INT | 100 | 0.01 | 16 | 0.25 | 1 | 700 | 0.02 | 16 | 1 | 0.25 |
| HIV_PROT | 100 | 0.02 | 16 | 1 | 0.7 | 700 | 0.01 | 16 | 0.25 | 0.25 |
| HPLC_LOGD | 1500 | 0.05 | 64 | 0.7 | 0.5 | 1500 | 0.05 | 64 | 1 | 0.25 |
| METAB | 1500 | 0.01 | 64 | 0.7 | 0.25 | 1500 | 0.05 | 64 | 1 | 0.5 |
| MK499 | 1500 | 0.02 | 128 | 0.7 | 0.25 | 1500 | 0.05 | 128 | 0.7 | 0.5 |
| NAV | 1500 | 0.02 | 256 | 1 | 0.25 | 1500 | 0.05 | 64 | 1 | 0.5 |
| NK1 | 700 | 0.02 | 32 | 1 | 0.25 | 700 | 0.02 | 16 | 0.25 | 0.25 |
| OX1 | 1500 | 0.05 | 32 | 0.5 | 0.25 | 1500 | 0.01 | 32 | 1 | 0.25 |
| OX2 | 1500 | 0.01 | 32 | 0.5 | 0.25 | 1500 | 0.02 | 32 | 0.7 | 0.25 |
| PAPP | 1500 | 0.05 | 64 | 1 | 0.25 | 1500 | 0.05 | 64 | 0.7 | 0.25 |
| PGP | 700 | 0.05 | 64 | 0.5 | 0.25 | 1500 | 0.02 | 32 | 0.7 | 0.25 |
| PPB | 1500 | 0.02 | 32 | 0.7 | 0.5 | 1500 | 0.02 | 32 | 0.5 | 0.25 |
| PXR | 1500 | 0.02 | 64 | 0.7 | 0.25 | 1500 | 0.05 | 128 | 0.7 | 0.25 |
| RAT_F | 350 | 0.02 | 32 | 0.7 | 0.25 | 700 | 0.02 | 32 | 0.7 | 0.5 |
| TDI | 100 | 0.02 | 64 | 0.7 | 0.7 | 700 | 0.01 | 64 | 0.5 | 0.7 |
| THROMBIN | 1500 | 0.02 | 32 | 0.25 | 0.25 | 1500 | 0.01 | 32 | 0.5 | 0.25 |

## 2. $R^2$ for STANDARD methods

| DATASET | R2 | TYPE |
|---|---|---|
| 2C8 | 0.158 | RF |
| 2C9BIG | 0.279 | RF |
| 2D6 | 0.13 | RF |
| 3A4 | 0.471 | RF |
| A-II | 0.805 | RF |
| BACE | 0.629 | RF |
| CAV | 0.399 | RF |
| CB1 | 0.292 | RF |
| CLINT | 0.393 | RF |
| DPP4 | 0.225 | RF |
| ERK2 | 0.257 | RF |
| FACTORXIA | 0.241 | RF |
| FASSIF | 0.294 | RF |
| HERG | 0.305 | RF |
| HERGBIG | 0.294 | RF |
| HIV_INT | 0.327 | RF |
| HIV_PROT | 0.545 | RF |
| HPLC_LOGD | 0.684 | RF |
| METAB | 0.631 | RF |
| NAV | 0.277 | RF |
| NK1 | 0.393 | RF |
| OX1 | 0.487 | RF |
| OX2 | 0.564 | RF |
| PAPP | 0.621 | RF |
| PGP | 0.556 | RF |
| PPB | 0.406 | RF |
| PXR | 0.333 | RF |
| RAT_F | 0.091 | RF |
| TDI | 0.385 | RF |
| THROMBIN | 0.242 | RF |
| 2C8 | 0.255 | DNN_STANDARD |
| 2C9BIG | 0.363 | DNN_STANDARD |
| 2D6 | 0.195 | DNN_STANDARD |
| 3A4 | 0.528 | DNN_STANDARD |
| A-II | 0.812 | DNN_STANDARD |
| BACE | 0.644 | DNN_STANDARD |
| CAV | 0.463 | DNN_STANDARD |
| CB1 | 0.321 | DNN_STANDARD |
| CLINT | 0.554 | DNN_STANDARD |
| DPP4 | 0.266 | DNN_STANDARD |
| ERK2 | 0.198 | DNN_STANDARD |
| FACTORXIA | 0.244 | DNN_STANDARD |
| FASSIF | 0.271 | DNN_STANDARD |
| HERG | 0.352 | DNN_STANDARD |
| HERGBIG | 0.367 | DNN_STANDARD |
| HIV_INT | 0.319 | DNN_STANDARD |
| HIV_PROT | 0.473 | DNN_STANDARD |
| HPLC_LOGD | 0.826 | DNN_STANDARD |
| METAB | 0.664 | DNN_STANDARD |
| NAV | 0.347 | DNN_STANDARD |
| NK1 | 0.422 | DNN_STANDARD |
| OX1 | 0.616 | DNN_STANDARD |
| OX2 | 0.599 | DNN_STANDARD |
| PAPP | 0.678 | DNN_STANDARD |
| PGP | 0.602 | DNN_STANDARD |
| PPB | 0.536 | DNN_STANDARD |
| PXR | 0.416 | DNN_STANDARD |
| RAT_F | 0.089 | DNN_STANDARD |
| TDI | 0.304 | DNN_STANDARD |
| THROMBIN | 0.313 | DNN_STANDARD |
| 2C8 | 0.207 | XGBOOST_TESTOPT |
| 2C9BIG | 0.344 | XGBOOST_TESTOPT |
| 2D6 | 0.19 | XGBOOST_TESTOPT |
| 3A4 | 0.517 | XGBOOST_TESTOPT |
| A-II | 0.802 | XGBOOST_TESTOPT |
| BACE | 0.656 | XGBOOST_TESTOPT |
| CAV | 0.459 | XGBOOST_TESTOPT |
| CB1 | 0.281 | XGBOOST_TESTOPT |
| CLINT | 0.474 | XGBOOST_TESTOPT |
| DPP4 | 0.23 | XGBOOST_TESTOPT |
| ERK2 | 0.287 | XGBOOST_TESTOPT |
| FACTORXIA | 0.386 | XGBOOST_TESTOPT |
| FASSIF | 0.318 | XGBOOST_TESTOPT |
| HERG | 0.355 | XGBOOST_TESTOPT |
| HERGBIG | 0.36 | XGBOOST_TESTOPT |
| HIV_INT | 0.298 | XGBOOST_TESTOPT |
| HIV_PROT | 0.587 | XGBOOST_TESTOPT |
| HPLC_LOGD | 0.804 | XGBOOST_TESTOPT |
| METAB | 0.625 | XGBOOST_TESTOPT |
| NAV | 0.33 | XGBOOST_TESTOPT |
| NK1 | 0.435 | XGBOOST_TESTOPT |
| OX1 | 0.578 | XGBOOST_TESTOPT |
| OX2 | 0.612 | XGBOOST_TESTOPT |
| PAPP | 0.646 | XGBOOST_TESTOPT |
| PGP | 0.606 | XGBOOST_TESTOPT |
| PPB | 0.538 | XGBOOST_TESTOPT |
| PXR | 0.391 | XGBOOST_TESTOPT |
| RAT_F | 0.125 | XGBOOST_TESTOPT |
| TDI | 0.38 | XGBOOST_TESTOPT |
| THROMBIN | 0.342 | XGBOOST_TESTOPT |
| 2C8 | 0.202 | XGBOOST_TRAINOPT |
| 2C9BIG | 0.344 | XGBOOST_TRAINOPT |
| 2D6 | 0.187 | XGBOOST_TRAINOPT |
| 3A4 | 0.515 | XGBOOST_TRAINOPT |
| A-II | 0.769 | XGBOOST_TRAINOPT |
| BACE | 0.651 | XGBOOST_TRAINOPT |
| CAV | 0.45 | XGBOOST_TRAINOPT |
| CB1 | 0.225 | XGBOOST_TRAINOPT |
| CLINT | 0.458 | XGBOOST_TRAINOPT |
| DPP4 | 0.218 | XGBOOST_TRAINOPT |
| ERK2 | 0.266 | XGBOOST_TRAINOPT |
| FACTORXIA | 0.271 | XGBOOST_TRAINOPT |
| FASSIF | 0.314 | XGBOOST_TRAINOPT |
| HERG | 0.352 | XGBOOST_TRAINOPT |
| HERGBIG | 0.355 | XGBOOST_TRAINOPT |
| HIV_INT | 0.273 | XGBOOST_TRAINOPT |
| HIV_PROT | 0.551 | XGBOOST_TRAINOPT |
| HPLC_LOGD | 0.804 | XGBOOST_TRAINOPT |
| METAB | 0.611 | XGBOOST_TRAINOPT |
| NAV | 0.328 | XGBOOST_TRAINOPT |
| NK1 | 0.433 | XGBOOST_TRAINOPT |
| OX1 | 0.553 | XGBOOST_TRAINOPT |
| OX2 | 0.589 | XGBOOST_TRAINOPT |
| PAPP | 0.639 | XGBOOST_TRAINOPT |
| PGP | 0.606 | XGBOOST_TRAINOPT |
| PPB | 0.526 | XGBOOST_TRAINOPT |
| PXR | 0.387 | XGBOOST_TRAINOPT |
| RAT_F | 0.103 | XGBOOST_TRAINOPT |
| TDI | 0.356 | XGBOOST_TRAINOPT |
| THROMBIN | 0.342 | XGBOOST_TRAINOPT |
| 2C8 | 0.181 | XGBOOST_STANDARD |
| 2C9BIG | 0.327 | XGBOOST_STANDARD |
| 2D6 | 0.178 | XGBOOST_STANDARD |
| 3A4 | 0.508 | XGBOOST_STANDARD |
| A-II | 0.778 | XGBOOST_STANDARD |
| BACE | 0.651 | XGBOOST_STANDARD |
| CAV | 0.452 | XGBOOST_STANDARD |
| CB1 | 0.253 | XGBOOST_STANDARD |
| CLINT | 0.461 | XGBOOST_STANDARD |
| DPP4 | 0.214 | XGBOOST_STANDARD |
| ERK2 | 0.253 | XGBOOST_STANDARD |
| FACTORXIA | 0.308 | XGBOOST_STANDARD |
| FASSIF | 0.315 | XGBOOST_STANDARD |
| HERG | 0.353 | XGBOOST_STANDARD |
| HERGBIG | 0.345 | XGBOOST_STANDARD |
| HIV_INT | 0.263 | XGBOOST_STANDARD |
| HIV_PROT | 0.502 | XGBOOST_STANDARD |
| HPLC_LOGD | 0.799 | XGBOOST_STANDARD |
| METAB | 0.621 | XGBOOST_STANDARD |
| NAV | 0.332 | XGBOOST_STANDARD |
| NK1 | 0.425 | XGBOOST_STANDARD |
| OX1 | 0.577 | XGBOOST_STANDARD |
| OX2 | 0.603 | XGBOOST_STANDARD |
| PAPP | 0.642 | XGBOOST_STANDARD |
| PGP | 0.592 | XGBOOST_STANDARD |
| PPB | 0.521 | XGBOOST_STANDARD |
| PXR | 0.389 | XGBOOST_STANDARD |
| RAT_F | 0.118 | XGBOOST_STANDARD |
| TDI | 0.359 | XGBOOST_STANDARD |
| THROMBIN | 0.288 | XGBOOST_STANDARD |
| 2C8 | 0.189 | LGB_ORIGINAL |
| 2C9BIG | 0.304 | LGB_ORIGINAL |
| 2D6 | 0.202 | LGB_ORIGINAL |
| 3A4 | 0.521 | LGB_ORIGINAL |
| A-II | 0.797 | LGB_ORIGINAL |
| BACE | 0.657 | LGB_ORIGINAL |
| CAV | 0.471 | LGB_ORIGINAL |
| CB1 | 0.351 | LGB_ORIGINAL |
| CLINT | 0.481 | LGB_ORIGINAL |
| DPP4 | 0.216 | LGB_ORIGINAL |
| ERK2 | 0.27 | LGB_ORIGINAL |
| FACTORXIA | 0.314 | LGB_ORIGINAL |
| FASSIF | 0.318 | LGB_ORIGINAL |
| HERG | 0.36 | LGB_ORIGINAL |
| HERGBIG | 0.32 | LGB_ORIGINAL |
| HIV_INT | 0.339 | LGB_ORIGINAL |
| HIV_PROT | 0.52 | LGB_ORIGINAL |
| HPLC_LOGD | 0.794 | LGB_ORIGINAL |
| METAB | 0.662 | LGB_ORIGINAL |
| NAV | 0.314 | LGB_ORIGINAL |
| NK1 | 0.435 | LGB_ORIGINAL |
| OX1 | 0.578 | LGB_ORIGINAL |
| OX2 | 0.614 | LGB_ORIGINAL |
| PAPP | 0.65 | LGB_ORIGINAL |
| PGP | 0.606 | LGB_ORIGINAL |
| PPB | 0.541 | LGB_ORIGINAL |
| PXR | 0.394 | LGB_ORIGINAL |
| RAT_F | 0.123 | LGB_ORIGINAL |
| TDI | 0.365 | LGB_ORIGINAL |
| THROMBIN | 0.321 | LGB_ORIGINAL |
| 2C8 | 0.226 | LGB_STANDARD |
| 2C9BIG | 0.348 | LGB_STANDARD |
| 2D6 | 0.21 | LGB_STANDARD |
| 3A4 | 0.545 | LGB_STANDARD |
| A-II | 0.789 | LGB_STANDARD |
| BACE | 0.639 | LGB_STANDARD |
| CAV | 0.475 | LGB_STANDARD |
| CB1 | 0.349 | LGB_STANDARD |
| CLINT | 0.513 | LGB_STANDARD |
| DPP4 | 0.218 | LGB_STANDARD |
| ERK2 | 0.24 | LGB_STANDARD |
| FACTORXIA | 0.262 | LGB_STANDARD |
| FASSIF | 0.33 | LGB_STANDARD |
| HERG | 0.373 | LGB_STANDARD |
| HERGBIG | 0.362 | LGB_STANDARD |
| HIV_INT | 0.311 | LGB_STANDARD |
| HIV_PROT | 0.485 | LGB_STANDARD |
| HPLC_LOGD | 0.821 | LGB_STANDARD |
| METAB | 0.655 | LGB_STANDARD |
| NAV | 0.342 | LGB_STANDARD |
| NK1 | 0.422 | LGB_STANDARD |
| OX1 | 0.59 | LGB_STANDARD |
| OX2 | 0.616 | LGB_STANDARD |
| PAPP | 0.659 | LGB_STANDARD |
| PGP | 0.603 | LGB_STANDARD |
| PPB | 0.535 | LGB_STANDARD |
| PXR | 0.414 | LGB_STANDARD |
| RAT_F | 0.109 | LGB_STANDARD |
| TDI | 0.358 | LGB_STANDARD |
| THROMBIN | 0.32 | LGB_STANDARD |
| 2C8 | 0.237 | LGB_TRAINOPT |
| 2C9BIG | 0.366 | LGB_TRAINOPT |
| 2D6 | 0.214 | LGB_TRAINOPT |
| 3A4 | 0.549 | LGB_TRAINOPT |
| A-II | 0.802 | LGB_TRAINOPT |
| BACE | 0.653 | LGB_TRAINOPT |
| CAV | 0.478 | LGB_TRAINOPT |
| CB1 | 0.35 | LGB_TRAINOPT |
| CLINT | 0.515 | LGB_TRAINOPT |
| DPP4 | 0.231 | LGB_TRAINOPT |
| ERK2 | 0.277 | LGB_TRAINOPT |
| FACTORXIA | 0.314 | LGB_TRAINOPT |
| FASSIF | 0.331 | LGB_TRAINOPT |
| HERG | 0.376 | LGB_TRAINOPT |
| HERGBIG | 0.362 | LGB_TRAINOPT |
| HIV_INT | 0.335 | LGB_TRAINOPT |
| HIV_PROT | 0.54 | LGB_TRAINOPT |
| HPLC_LOGD | 0.82 | LGB_TRAINOPT |
| METAB | 0.651 | LGB_TRAINOPT |
| NAV | 0.333 | LGB_TRAINOPT |
| NK1 | 0.442 | LGB_TRAINOPT |
| OX1 | 0.603 | LGB_TRAINOPT |
| OX2 | 0.627 | LGB_TRAINOPT |
| PAPP | 0.661 | LGB_TRAINOPT |
| PGP | 0.606 | LGB_TRAINOPT |
| PPB | 0.552 | LGB_TRAINOPT |
| PXR | 0.422 | LGB_TRAINOPT |
| RAT_F | 0.125 | LGB_TRAINOPT |
| TDI | 0.378 | LGB_TRAINOPT |
| THROMBIN | 0.335 | LGB_TRAINOPT |

## 3. Timing for all methods

| Data set | Ntraining | TYPE | TOTAL TIME (HOURS) |
|---|---|---|---|
| 2C8 | 22500 | RF | 4 |
| 2C9BIG | 142000 | RF | 224 |
| 2D6 | 37500 | RF | 19 |
| 3A4 | 37499 | RF | 15 |
| A-II | 2072 | RF | 0.1 |
| BACE | 13101 | RF | 2 |
| CAV | 37500 | RF | 17 |
| CB1 | 8730 | RF | 1.1 |
| CLINT | 17469 | RF | 3.7 |
| DPP4 | 6150 | RF | 0.9 |
| ERK2 | 9632 | RF | 2.3 |
| FACTORXIA | 7149 | RF | 0.9 |
| FASSIF | 67100 | RF | 33.3 |
| HERG | 37473 | RF | 9.6 |
| HERGBIG | 238000 | RF | 494 |
| HIV_INT | 1815 | RF | 0.1 |
| HIV_PROT | 3233 | RF | 0.4 |
| HPLC_LOGD | 37500 | RF | 16 |
| METAB | 1569 | RF | 0.2 |
| NAV | 34682 | RF | 10.1 |
| NK1 | 10050 | RF | 1.3 |
| OX1 | 5351 | RF | 0.5 |
| OX2 | 11156 | RF | 1.8 |
| PAPP | 23204 | RF | 5.2 |
| PGP | 6450 | RF | 0.5 |
| PPB | 8716 | RF | 0.8 |
| PXR | 37499 | RF | 19.8 |
| RAT_F | 6109 | RF | 0.5 |
| TDI | 4169 | RF | 0.4 |
| THROMBIN | 5100 | RF | 0.4 |
| 2C8 | 22500 | DNN_STANDARD | 20 |
| 2C9BIG | 142000 | DNN_STANDARD | 167 |
| 2D6 | 37500 | DNN_STANDARD | 37.5 |
| 3A4 | 37499 | DNN_STANDARD | 38 |
| A-II | 2072 | DNN_STANDARD | 1.5 |
| BACE | 13101 | DNN_STANDARD | 9.83 |
| CAV | 37500 | DNN_STANDARD | 36 |
| CB1 | 8730 | DNN_STANDARD | 6.5 |
| CLINT | 17469 | DNN_STANDARD | 13.84 |
| DPP4 | 6150 | DNN_STANDARD | 4.33 |
| ERK2 | 9632 | DNN_STANDARD | 7.67 |
| FACTORXIA | 7149 | DNN_STANDARD | 5.33 |
| FASSIF | 67100 | DNN_STANDARD | 68 |
| HERG | 37473 | DNN_STANDARD | 36.67 |
| HERGBIG | 238000 | DNN_STANDARD | 292.5 |
| HIV_INT | 1815 | DNN_STANDARD | 1.17 |
| HIV_PROT | 3233 | DNN_STANDARD | 2.5 |
| HPLC_LOGD | 37500 | DNN_STANDARD | 36 |
| METAB | 1569 | DNN_STANDARD | 1 |
| NAV | 34682 | DNN_STANDARD | 31.83 |
| NK1 | 10050 | DNN_STANDARD | 7.33 |
| OX1 | 5351 | DNN_STANDARD | 3.5 |
| OX2 | 11156 | DNN_STANDARD | 8.17 |
| PAPP | 23204 | DNN_STANDARD | 20 |
| PGP | 6450 | DNN_STANDARD | 4.17 |
| PPB | 8716 | DNN_STANDARD | 6 |
| PXR | 37499 | DNN_STANDARD | 40.1 |
| RAT_F | 6109 | DNN_STANDARD | 4.67 |
| TDI | 4169 | DNN_STANDARD | 3.16 |
| THROMBIN | 5100 | DNN_STANDARD | 3.66 |
| 2C8 | 22500 | DNN_QUICK | 2.66 |
| 2C9BIG | 142000 | DNN_QUICK | 23 |
| 2D6 | 37500 | DNN_QUICK | 5 |
| 3A4 | 37499 | DNN_QUICK | 5 |
| A-II | 2072 | DNN_QUICK | 0.17 |
| BACE | 13101 | DNN_QUICK | 1.33 |
| CAV | 37500 | DNN_QUICK | 4.67 |
| CB1 | 8730 | DNN_QUICK | 0.66 |
| CLINT | 17469 | DNN_QUICK | 1.67 |
| DPP4 | 6150 | DNN_QUICK | 0.5 |
| ERK2 | 9632 | DNN_QUICK | 0.83 |
| FACTORXIA | 7149 | DNN_QUICK | 0.67 |
| FASSIF | 67100 | DNN_QUICK | 9 |
| HERG | 37473 | DNN_QUICK | 4.83 |
| HERGBIG | 238000 | DNN_QUICK | 40 |
| HIV_INT | 1815 | DNN_QUICK | 0.16 |
| HIV_PROT | 3233 | DNN_QUICK | 0.34 |
| HPLC_LOGD | 37500 | DNN_QUICK | 4.83 |
| METAB | 1569 | DNN_QUICK | 0 |
| NAV | 34682 | DNN_QUICK | 3.83 |
| NK1 | 10050 | DNN_QUICK | 0.84 |
| OX1 | 5351 | DNN_QUICK | 0.5 |
| OX2 | 11156 | DNN_QUICK | 1 |
| PAPP | 23204 | DNN_QUICK | 2.5 |
| PGP | 6450 | DNN_QUICK | 0.5 |
| PPB | 8716 | DNN_QUICK | 0.67 |
| PXR | 37499 | DNN_QUICK | 4.83 |
| RAT_F | 6109 | DNN_QUICK | 0.5 |
| TDI | 4169 | DNN_QUICK | 0.34 |
| THROMBIN | 5100 | DNN_QUICK | 0.5 |
| 2C8 | 22500 | XGBOOST_STANDARD | 0.15 |
| 2C9BIG | 142000 | XGBOOST_STANDARD | 1.2 |
| 2D6 | 37500 | XGBOOST_STANDARD | 0.28 |
| 3A4 | 37499 | XGBOOST_STANDARD | 0.266 |
| A-II | 2072 | XGBOOST_STANDARD | 0.017 |
| BACE | 13101 | XGBOOST_STANDARD | 0.05 |
| CAV | 37500 | XGBOOST_STANDARD | 0.25 |
| CB1 | 8730 | XGBOOST_STANDARD | 0.05 |
| CLINT | 17469 | XGBOOST_STANDARD | 0.084 |
| DPP4 | 6150 | XGBOOST_STANDARD | 0.034 |
| ERK2 | 9632 | XGBOOST_STANDARD | 0.05 |
| FACTORXIA | 7149 | XGBOOST_STANDARD | 0.034 |
| FASSIF | 67100 | XGBOOST_STANDARD | 0.5 |
| HERG | 37473 | XGBOOST_STANDARD | 0.266 |
| HERGBIG | 238000 | XGBOOST_STANDARD | 2.05 |
| HIV_INT | 1815 | XGBOOST_STANDARD | 0.01 |
| HIV_PROT | 3233 | XGBOOST_STANDARD | 0.017 |
| HPLC_LOGD | 37500 | XGBOOST_STANDARD | 0.25 |
| METAB | 1569 | XGBOOST_STANDARD | 0.017 |
| NAV | 34682 | XGBOOST_STANDARD | 0.217 |
| NK1 | 10050 | XGBOOST_STANDARD | 0.05 |
| OX1 | 5351 | XGBOOST_STANDARD | 0.033 |
| OX2 | 11156 | XGBOOST_STANDARD | 0.05 |
| PAPP | 23204 | XGBOOST_STANDARD | 0.133 |
| PGP | 6450 | XGBOOST_STANDARD | 0.033 |
| PPB | 8716 | XGBOOST_STANDARD | 0.033 |
| PXR | 37499 | XGBOOST_STANDARD | 0.25 |
| RAT_F | 6109 | XGBOOST_STANDARD | 0.033 |
| TDI | 4169 | XGBOOST_STANDARD | 0.017 |
| THROMBIN | 5100 | XGBOOST_STANDARD | 0.033 |
| 2C8 | 22500 | LGB_ORIGINAL | 0.027 |
| 2C9BIG | 142000 | LGB_ORIGINAL | 0.481 |
| 2D6 | 37500 | LGB_ORIGINAL | 0.053 |
| 3A4 | 37499 | LGB_ORIGINAL | 0.051 |
| A-II | 2072 | LGB_ORIGINAL | 0.004 |
| BACE | 13101 | LGB_ORIGINAL | 0.014 |
| CAV | 37500 | LGB_ORIGINAL | 0.047 |
| CB1 | 8730 | LGB_ORIGINAL | 0.009 |
| CLINT | 17469 | LGB_ORIGINAL | 0.018 |
| DPP4 | 6150 | LGB_ORIGINAL | 0.007 |
| ERK2 | 9632 | LGB_ORIGINAL | 0.017 |
| FACTORXIA | 7149 | LGB_ORIGINAL | 0.008 |
| FASSIF | 67100 | LGB_ORIGINAL | 0.1 |
| HERG | 37473 | LGB_ORIGINAL | 0.051 |
| HERGBIG | 238000 | LGB_ORIGINAL | 0.51 |
| HIV_INT | 1815 | LGB_ORIGINAL | 0.005 |
| HIV_PROT | 3233 | LGB_ORIGINAL | 0.005 |
| HPLC_LOGD | 37500 | LGB_ORIGINAL | 0.048 |
| METAB | 1569 | LGB_ORIGINAL | 0.003 |
| NAV | 34682 | LGB_ORIGINAL | 0.041 |
| NK1 | 10050 | LGB_ORIGINAL | 0.01 |
| OX1 | 5351 | LGB_ORIGINAL | 0.006 |
| OX2 | 11156 | LGB_ORIGINAL | 0.011 |
| PAPP | 23204 | LGB_ORIGINAL | 0.027 |
| PGP | 6450 | LGB_ORIGINAL | 0.007 |
| PPB | 8716 | LGB_ORIGINAL | 0.009 |
| PXR | 37499 | LGB_ORIGINAL | 0.049 |
| RAT_F | 6109 | LGB_ORIGINAL | 0.008 |
| TDI | 4169 | LGB_ORIGINAL | 0.006 |
| THROMBIN | 5100 | LGB_ORIGINAL | 0.006 |
| 2C8 | 22500 | LGB_STANDARD | 0.045 |
| 2C9BIG | 142000 | LGB_STANDARD | 0.293 |
| 2D6 | 37500 | LGB_STANDARD | 0.08 |
| 3A4 | 37499 | LGB_STANDARD | 0.064 |
| A-II | 2072 | LGB_STANDARD | 0.007 |
| BACE | 13101 | LGB_STANDARD | 0.024 |
| CAV | 37500 | LGB_STANDARD | 0.074 |
| CB1 | 8730 | LGB_STANDARD | 0.012 |
| CLINT | 17469 | LGB_STANDARD | 0.032 |
| DPP4 | 6150 | LGB_STANDARD | 0.008 |
| ERK2 | 9632 | LGB_STANDARD | 0.019 |
| FACTORXIA | 7149 | LGB_STANDARD | 0.017 |
| FASSIF | 67100 | LGB_STANDARD | 0.14 |
| HERG | 37473 | LGB_STANDARD | 0.077 |
| HERGBIG | 238000 | LGB_STANDARD | 0.574 |
| HIV_INT | 1815 | LGB_STANDARD | 0.003 |
| HIV_PROT | 3233 | LGB_STANDARD | 0.006 |
| HPLC_LOGD | 37500 | LGB_STANDARD | 0.06 |
| METAB | 1569 | LGB_STANDARD | 0.003 |
| NAV | 34682 | LGB_STANDARD | 0.065 |
| NK1 | 10050 | LGB_STANDARD | 0.013 |
| OX1 | 5351 | LGB_STANDARD | 0.007 |
| OX2 | 11156 | LGB_STANDARD | 0.014 |
| PAPP | 23204 | LGB_STANDARD | 0.045 |
| PGP | 6450 | LGB_STANDARD | 0.008 |
| PPB | 8716 | LGB_STANDARD | 0.011 |
| PXR | 37499 | LGB_STANDARD | 0.074 |
| RAT_F | 6109 | LGB_STANDARD | 0.009 |
| TDI | 4169 | LGB_STANDARD | 0.007 |
| THROMBIN | 5100 | LGB_STANDARD | 0.008 |

4. File size for all methods

| Data set | Ntraining | METHOD | FILE_SIZE(MEGABYTES) |
|---|---|---|---|
| 2C8 | 22500 | DNN_QUICK | 32.78 |
| 2C9BIG | 142000 | DNN_QUICK | 46.833 |
| 2D6 | 37500 | DNN_QUICK | 38.825 |
| 3A4 | 37499 | DNN_QUICK | 38.8531 |
| A-II | 2072 | DNN_QUICK | 21.1875 |
| BACE | 13101 | DNN_QUICK | 25.8076 |
| CAV | 37500 | DNN_QUICK | 36.1236 |
| CB1 | 8730 | DNN_QUICK | 24.3144 |
| CLINT | 17469 | DNN_QUICK | 26.9597 |
| DPP4 | 6150 | DNN_QUICK | 22.187 |
| ERK2 | 9632 | DNN_QUICK | 27.341 |
| FACTORXIA | 7149 | DNN_QUICK | 24.6837 |
| FASSIF | 67100 | DNN_QUICK | 38.9093 |
| HERG | 37473 | DNN_QUICK | 38.3915 |
| HERGBIG | 238000 | DNN_QUICK | 49.3819 |
| HIV_INT | 1815 | DNN_QUICK | 18.8193 |
| HIV_PROT | 3233 | DNN_QUICK | 25.1012 |
| HPLC_LOGD | 37500 | DNN_QUICK | 36.6294 |
| METAB | 1569 | DNN_QUICK | 19.5659 |
| NAV | 34682 | DNN_QUICK | 33.936 |
| NK1 | 10050 | DNN_QUICK | 24.4629 |
| OX1 | 5351 | DNN_QUICK | 20.4851 |
| OX2 | 11156 | DNN_QUICK | 23.9411 |
| PAPP | 23204 | DNN_QUICK | 31.4834 |
| PGP | 6450 | DNN_QUICK | 21.0069 |
| PPB | 8716 | DNN_QUICK | 22.0505 |
| PXR | 37499 | DNN_QUICK | 37.7613 |
| RAT_F | 6109 | DNN_QUICK | 24.194 |
| TDI | 4169 | DNN_QUICK | 24.9446 |
| THROMBIN | 5100 | DNN_QUICK | 23.2186 |
| 2C8 | 22500 | DNN_STANDARD | 166.802 |
| 2C9BIG | 142000 | DNN_STANDARD | 222.867 |
| 2D6 | 37500 | DNN_STANDARD | 190.919 |
| 3A4 | 37499 | DNN_STANDARD | 191.031 |
| A-II | 2072 | DNN_STANDARD | 120.554 |
| BACE | 13101 | DNN_STANDARD | 138.986 |
| CAV | 37500 | DNN_STANDARD | 180.142 |
| CB1 | 8730 | DNN_STANDARD | 133.029 |
| CLINT | 17469 | DNN_STANDARD | 143.582 |
| DPP4 | 6150 | DNN_STANDARD | 124.541 |
| ERK2 | 9632 | DNN_STANDARD | 145.103 |
| FACTORXIA | 7149 | DNN_STANDARD | 134.502 |
| FASSIF | 67100 | DNN_STANDARD | 191.256 |
| HERG | 37473 | DNN_STANDARD | 189.19 |
| HERGBIG | 238000 | DNN_STANDARD | 233.036 |
| HIV_INT | 1815 | DNN_STANDARD | 111.105 |
| HIV_PROT | 3233 | DNN_STANDARD | 136.167 |
| HPLC_LOGD | 37500 | DNN_STANDARD | 182.16 |
| METAB | 1569 | DNN_STANDARD | 114.084 |
| NAV | 34682 | DNN_STANDARD | 171.414 |
| NK1 | 10050 | DNN_STANDARD | 133.621 |
| OX1 | 5351 | DNN_STANDARD | 117.751 |
| OX2 | 11156 | DNN_STANDARD | 131.539 |
| PAPP | 23204 | DNN_STANDARD | 161.63 |
| PGP | 6450 | DNN_STANDARD | 119.833 |
| PPB | 8716 | DNN_STANDARD | 123.997 |
| PXR | 37499 | DNN_STANDARD | 186.676 |
| RAT_F | 6109 | DNN_STANDARD | 132.548 |
| TDI | 4169 | DNN_STANDARD | 135.543 |
| THROMBIN | 5100 | DNN_STANDARD | 128.657 |
| 2C8 | 22500 | LGB_ORIGINAL | 4.33 |
| 2C9BIG | 142000 | LGB_ORIGINAL | 4.47 |
| 2D6 | 37500 | LGB_ORIGINAL | 4.35 |
| 3A4 | 37499 | LGB_ORIGINAL | 4.35 |
| A-II | 2072 | LGB_ORIGINAL | 4.17 |
| BACE | 13101 | LGB_ORIGINAL | 4.28 |
| CAV | 37500 | LGB_ORIGINAL | 4.38 |
| CB1 | 8730 | LGB_ORIGINAL | 4.25 |
| CLINT | 17469 | LGB_ORIGINAL | 4.32 |
| DPP4 | 6150 | LGB_ORIGINAL | 4.24 |
| ERK2 | 9632 | LGB_ORIGINAL | 4.27 |
| FACTORXIA | 7149 | LGB_ORIGINAL | 4.26 |
| FASSIF | 67100 | LGB_ORIGINAL | 4.43 |
| HERG | 37473 | LGB_ORIGINAL | 4.38 |
| HERGBIG | 238000 | LGB_ORIGINAL | 4.38 |
| HIV_INT | 1815 | LGB_ORIGINAL | 4.21 |
| HIV_PROT | 3233 | LGB_ORIGINAL | 4.21 |
| HPLC_LOGD | 37500 | LGB_ORIGINAL | 4.39 |
| METAB | 1569 | LGB_ORIGINAL | 4.06 |
| NAV | 34682 | LGB_ORIGINAL | 4.36 |
| NK1 | 10050 | LGB_ORIGINAL | 4.23 |
| OX1 | 5351 | LGB_ORIGINAL | 4.23 |
| OX2 | 11156 | LGB_ORIGINAL | 4.24 |
| PAPP | 23204 | LGB_ORIGINAL | 4.39 |
| PGP | 6450 | LGB_ORIGINAL | 4.29 |
| PPB | 8716 | LGB_ORIGINAL | 4.28 |
| PXR | 37499 | LGB_ORIGINAL | 4.21 |
| RAT_F | 6109 | LGB_ORIGINAL | 4.3 |
| TDI | 4169 | LGB_ORIGINAL | 4.3 |
| THROMBIN | 5100 | LGB_ORIGINAL | 4.27 |
| 2C8 | 22500 | LGB_STANDARD | 7.5 |
| 2C9BIG | 142000 | LGB_STANDARD | 7.6 |
| 2D6 | 37500 | LGB_STANDARD | 7.6 |
| 3A4 | 37499 | LGB_STANDARD | 7.5 |
| A-II | 2072 | LGB_STANDARD | 5.8 |
| BACE | 13101 | LGB_STANDARD | 7.5 |
| CAV | 37500 | LGB_STANDARD | 7.6 |
| CB1 | 8730 | LGB_STANDARD | 7.4 |
| CLINT | 17469 | LGB_STANDARD | 7.5 |
| DPP4 | 6150 | LGB_STANDARD | 7.4 |
| ERK2 | 9632 | LGB_STANDARD | 7.5 |
| FACTORXIA | 7149 | LGB_STANDARD | 7.5 |
| FASSIF | 67100 | LGB_STANDARD | 7.6 |
| HERG | 37473 | LGB_STANDARD | 7.5 |
| HERGBIG | 238000 | LGB_STANDARD | 7.7 |
| HIV_INT | 1815 | LGB_STANDARD | 5.3 |
| HIV_PROT | 3233 | LGB_STANDARD | 7.4 |
| HPLC_LOGD | 37500 | LGB_STANDARD | 7.6 |
| METAB | 1569 | LGB_STANDARD | 4.3 |
| NAV | 34682 | LGB_STANDARD | 7.6 |
| NK1 | 10050 | LGB_STANDARD | 7.4 |
| OX1 | 5351 | LGB_STANDARD | 7.4 |
| OX2 | 11156 | LGB_STANDARD | 7.4 |
| PAPP | 23204 | LGB_STANDARD | 7.6 |
| PGP | 6450 | LGB_STANDARD | 7.6 |
| PPB | 8716 | LGB_STANDARD | 7.5 |
| PXR | 37499 | LGB_STANDARD | 7.2 |
| RAT_F | 6109 | LGB_STANDARD | 7.5 |
| TDI | 4169 | LGB_STANDARD | 7.6 |
| THROMBIN | 5100 | LGB_STANDARD | 7.4 |
| 2C8 | 22500 | RF | 21.6032 |
| 2C9BIG | 142000 | RF | 135.2 |
| 2D6 | 37500 | RF | 35.6864 |
| 3A4 | 37499 | RF | 35.7536 |
| A-II | 2072 | RF | 1.9856 |
| BACE | 13101 | RF | 12.5792 |
| CAV | 37500 | RF | 35.5136 |
| CB1 | 8730 | RF | 8.3696 |
| CLINT | 17469 | RF | 16.7552 |
| DPP4 | 6150 | RF | 5.9024 |
| ERK2 | 9632 | RF | 9.2432 |
| FACTORXIA | 7149 | RF | 6.8576 |
| FASSIF | 67100 | RF | 64.3664 |
| HERG | 37473 | RF | 35.5856 |
| HERGBIG | 238000 | RF | 228.483 |
| HIV_INT | 1815 | RF | 1.7456 |
| HIV_PROT | 3233 | RF | 3.0848 |
| HPLC_LOGD | 37500 | RF | 35.8928 |
| METAB | 1569 | RF | 1.5056 |
| NAV | 34682 | RF | 32.8352 |
| NK1 | 10050 | RF | 9.5696 |
| OX1 | 5351 | RF | 5.1392 |
| OX2 | 11156 | RF | 10.7072 |
| PAPP | 23204 | RF | 21.872 |
| PGP | 6450 | RF | 6.1424 |
| PPB | 8716 | RF | 8.3072 |
| PXR | 37499 | RF | 35.7152 |
| RAT_F | 6109 | RF | 5.864 |
| TDI | 4169 | RF | 4.0016 |
| THROMBIN | 5100 | RF | 4.856 |
| 2C8 | 22500 | XGBOOST_STANDARD | 3.63065 |
| 2C9BIG | 142000 | XGBOOST_STANDARD | 5.53562 |
| 2D6 | 37500 | XGBOOST_STANDARD | 3.56966 |
| 3A4 | 37499 | XGBOOST_STANDARD | 3.92686 |
| A-II | 2072 | XGBOOST_STANDARD | 2.25242 |
| BACE | 13101 | XGBOOST_STANDARD | 3.49824 |
| CAV | 37500 | XGBOOST_STANDARD | 4.2343 |
| CB1 | 8730 | XGBOOST_STANDARD | 3.07639 |
| CLINT | 17469 | XGBOOST_STANDARD | 3.97027 |
| DPP4 | 6150 | XGBOOST_STANDARD | 2.9671 |
| ERK2 | 9632 | XGBOOST_STANDARD | 2.82245 |
| FACTORXIA | 7149 | XGBOOST_STANDARD | 2.70401 |
| FASSIF | 67100 | XGBOOST_STANDARD | 4.16726 |
| HERG | 37473 | XGBOOST_STANDARD | 4.7249 |
| HERGBIG | 238000 | XGBOOST_STANDARD | 6.5663 |
| HIV_INT | 1815 | XGBOOST_STANDARD | 2.23853 |
| HIV_PROT | 3233 | XGBOOST_STANDARD | 2.79185 |
| HPLC_LOGD | 37500 | XGBOOST_STANDARD | 5.21306 |
| METAB | 1569 | XGBOOST_STANDARD | 2.21297 |
| NAV | 34682 | XGBOOST_STANDARD | 3.78602 |
| NK1 | 10050 | XGBOOST_STANDARD | 3.28231 |
| OX1 | 5351 | XGBOOST_STANDARD | 3.17244 |
| OX2 | 11156 | XGBOOST_STANDARD | 3.6773 |
| PAPP | 23204 | XGBOOST_STANDARD | 3.39024 |
| PGP | 6450 | XGBOOST_STANDARD | 3.39031 |
| PPB | 8716 | XGBOOST_STANDARD | 3.74311 |
| PXR | 37499 | XGBOOST_STANDARD | 4.34647 |
| RAT_F | 6109 | XGBOOST_STANDARD | 2.79703 |
| TDI | 4169 | XGBOOST_STANDARD | 2.86529 |
| THROMBIN | 5100 | XGBOOST_STANDARD | 3.02815 |